\documentclass[conference]{IEEEtran}
\IEEEoverridecommandlockouts

\usepackage{graphicx}
\usepackage{eqparbox}

\usepackage{subcaption}
\usepackage{xcolor}
\usepackage{multirow}

\usepackage{cite}
\usepackage{amsmath,amssymb,amsfonts}
\usepackage{algorithmic}

\usepackage{textcomp}
\usepackage{setspace}
\usepackage[hyphens]{url}
\usepackage{hyperref}
\usepackage[hyphenbreaks]{breakurl}
\def\BibTeX{{\rm B\kern-.05em{\sc i\kern-.025em b}\kern-.08em
    T\kern-.1667em\lower.7ex\hbox{E}\kern-.125emX}}

\makeatletter
\def\@IEEEpubidpullup{8\baselineskip}
\makeatother
\begin{document}
\IEEEoverridecommandlockouts
\IEEEpubid{
\parbox{\columnwidth}{\vspace{-4\baselineskip}Permission to make digital or hard copies of all or part of this work for personal or classroom use is granted without fee provided that copies are not made or distributed for profit or commercial advantage and that copies bear this notice and the full citation on the first page. Copyrights for components of this work owned by others than ACM must be honored. Abstracting with credit is permitted. To copy otherwise, or republish, to post on servers or to redistribute to lists, requires prior specific permission and/or a fee. Request permissions from  \href{mailto:permissions@acm.org}{Permissions@acm.org}.\hfill\vspace{-0.8\baselineskip}\\
\begin{spacing}{1.2}
\small\textit{ASONAM '21}, November 8-11, 2021, Virtual Event, Netherlands \\
\copyright\space2021 Association for Computing Machinery. \\
ACM ISBN 978-1-4503-9128-3/21/11...\$15.00 \\
\url{https://doi.org/10.1145/3487351.3492718}
\end{spacing}
\hfill}
\hspace{0.9\columnsep}\makebox[\columnwidth]{\hfill}}
\IEEEpubidadjcol
\bstctlcite{IEEEexample:BSTcontrol}


\title{Analyzing Behavioral Changes of Twitter Users After Exposure to Misinformation\\
}

\author{\IEEEauthorblockN{Yichen Wang,
Richard Han,
Tamara Lehman,
Qin Lv, and
Shivakant Mishra}
\IEEEauthorblockA{\textit{University of Colorado Boulder, Boulder, CO, USA}}
\IEEEauthorblockA{\{yichen.wang, richard.han, tamara.lehman, qin.lv, shivakaht.mishra\}@colorado.edu}}

\maketitle

\begin{abstract}
Social media platforms have been exploited to disseminate misinformation in recent years. The widespread online misinformation has been shown to affect users' beliefs and is connected to social impact such as polarization. In this work, we focus on misinformation's impact on specific user behavior and aim to understand whether general Twitter users changed their behavior after being exposed to misinformation. We compare the before and after behavior of exposed users to determine whether the frequency of the tweets they posted, or the sentiment of their tweets underwent any significant change. Our results indicate that users overall exhibited statistically significant changes in behavior across some of these metrics. Through language distance analysis, we show that exposed users were already different from baseline users before the exposure. We also study the characteristics of two specific user groups, multi-exposure and extreme change groups, which were potentially highly impacted. Finally, we study if the changes in the behavior of the users after exposure to misinformation tweets vary based on the number of their followers or the number of followers of the tweet authors, and find that their behavioral changes are all similar.
\end{abstract}

\begin{IEEEkeywords}
Misinformation, Fake News, Twitter, User Behavior
\end{IEEEkeywords}

\section{Introduction}
Online social media has become increasingly popular in recent years and has been used to disseminate misinformation by users, sometimes intentionally, resulting in detrimental effects on our society. For example, some participants in the 2021 United States (US) Capitol riot said they were driven by online misinformation and conspiracy theories~\cite{riot1,riot2}. As another example, misinformation is still driving people's vaccine hesitancy, especially during the COVID-19 pandemic~\cite{vacMis}. The spread of misinformation is a real threat to our society, as it can disrupt the public trust of legitimate news sources and undermine the political spectrum.

To combat misinformation, researchers have focused on two aspects: detecting misinformation and understanding its impact. To detect misinformation, researchers have built models making use of various information including content style, user profile and social context~\cite{Perez-Rosas2017-tt,shu2018understanding,Shu2019-ig}. To make it more amenable to the masses, academics have also proposed mechanisms to automate the fact-checking process~\cite{hassan2017toward,Ciampaglia2015-nu}. 

To study misinformation's impact, researchers have investigated the spread pattern of misinformation \cite{Del_Vicario2016-rb,Vosoughi2018-qt}, its negative effect on users' beliefs \cite{Nyhan2010-sf,Bessi2015-cc,Mocanu2015-hk},  and its correlation with some social phenomena such as echo chambers and polarization \cite{Ribeiro2017-uk}. However, prior work has focused more on the misinformation's general social effect, and very little work has been done to examine what and how specific user behavior is affected. We argue that it is crucial to study the details of specific behavioral changes after being exposed to misinformation. It can help us understand the process of how users succumb to misinformation and get affected negatively by exposure to misinformation. It can also help us identify specific user groups who are more likely to be vulnerable to misinformation and potentially even be radicalized. Some previous work studied the impact of COVID related misinformation on users' vaccine intent \cite{Loomba2021-ai}, but they only focus on this specific type of misinformation. Limited work has focused on misinformation with broader topics and users' individual behavioral change.

In this paper, we have conducted a large-scale, quantitative analysis of Twitter user behavior after exposure to a known piece of misinformation. A user is considered to have been exposed to misinformation if he/she replies to a tweet carrying misinformation ({\it misinformation tweet}). We believe that the action of replying to a misinformation tweet is a much stronger indication of a user being exposed to misinformation and being influenced by it than other actions such as reading or liking a tweet. Specifically, to understand users' behavioral change, we seek to answer the following research questions (RQs):
\begin{itemize}
    \item RQ1: Do the users who reply to misinformation tweets exhibit a change in their behavior after the exposure?
    \item RQ2: Does the change in behavior of users who reply to {\it multiple} misinformation tweets differ from other users? 
    \item RQ3: What are the characteristics of the users who undergo extreme behavioral change after being exposed to misinformation tweets?
    \item RQ4: Does the changes in user behavior of the users after being exposed to misinformation tweets vary based on the number of their followers or the number of followers of the tweet authors? 
\end{itemize}

To identify misinformation tweets, we first obtained fact-checked misinformation excerpts from the well-known fact-checking website PolitiFact, and then queried Twitter to collect those tweets that contained these misinformation excerpts. Next, we collected the identities of all the users who replied to these misinformation tweets (named ``target group" in the remaining part of this paper). To establish that any change in user behavior we observe is potentially due to exposure to misinformation tweets, we also built a user-controlled baseline group (named ``baseline 1" in the remaining part of this paper) and an entity-matched baseline group (named ``baseline 2" in the remaining part of this paper) for comparison. To identify whether there were significant changes in the tweeting behavior before and after exposure to misinformation, we selected objective behavioral metrics such as mean tweet count, mean sentiment score of tweets, and language usage distance. Then, we analyzed these behavioral metrics before and after exposure in both short-term (twenty-four hours before and after) and long-term (six months before and after). Overall, this paper makes the following contributions:
\begin{itemize}
    \item We introduce a dataset containing 372 misinformation tweets along with 21,071 users who replied to them and their tweets from six months before until six months after their reply.
    \item We reveal evidence of statistically significant changes in the number and frequency of tweets that users send after being exposed to misinformation tweets, both in the short and long terms. We do not observe such changes in the baseline user groups, indicating that there is a positive correlation between increase in count/frequency of tweets and exposure to misinformation tweets.
    \item We do not find any significant change in user's overall tweet sentiments after being exposed to misinformation tweets. Using language distance analysis, we show that baseline users already had different language characteristics than the exposed users even before the exposure.
    \item We investigate the group of users exposed to multiple misinformation tweets, i.e. they replied to more than one misinformation tweet. We find that these users’ behavioral change is less significant when compared with the users exposed to a single misinformation tweet, and that these users were already on a high activity level before the exposures.
    \item We examine the group of users who had extreme behavioral changes and find that users with extreme changes of tweet count do not overlap with the group of users with extreme change of sentiment. Further, users with extreme changes in the short term do not overlap with the users with extreme changes in the long-term.
    \item We find that exposed users with high and low follower counts exhibit similar behavioral change (significantly increase tweeting frequency). Further, we find users exposed to misinformation tweets authored by users with high and low follower counts also undergo similar behavioral change, while the users exposed to low-follower-count-authored tweets generally post more tweets.
\end{itemize}

The intent of this paper is to identify and quantify correlation between exposure to Twitter misinformation and changed behavior in the exposed users where it exists, and not to establish causality, i.e., the paper does not claim that the changed behavior is caused by exposure to misinformation. Establishing causality would require further research. The Discussion section of the paper describes this in more detail.

The organization of the rest of our paper is as follows. In Section II, we describe the background and related work on misinformation on social media. In Section III, we present and explain the methodologies used to create the dataset and the user behavioral features under study. In Section IV, we present the results of the analysis of the research questions. Finally, in Section V, we conclude the work and discuss its implications, limitations, and possible future directions.

\section{Related Work}
Researchers have studied users and content on online social platforms extensively~\cite{benevenuto2009characterizing,jin2013understanding,wang2020jump}, and it has been shown that online social media has become a major source of misinformation~\cite{picchi_2018,alba_2020,Tallal_Javed2020-cp}. A significant body of work has investigated the spread and detection of misinformation. Mustafaraj et al. described the spread process of fake news~\cite{mustafaraj2017fake}. The diffusion process is also modeled by Tambusc et al.~\cite{tambuscio2015fact}. Making use of abundant data from a social network, Vosoughi et al. studied the spread pattern of fake news on Twitter from 2006--2017 and found that fake news spread farther, faster, deeper, and more broadly than true news~\cite{Vosoughi2018-qt}. Vicario et al. studied the conspiracy news spreading on Facebook and found selective exposure is the primary driver of the diffusion~\cite{Del_Vicario2016-rb}. To combat misinformation, scientists have studied and used a wide range of detection techniques. Journalists and investigators have built many manual fact-checking websites\footnote{https://www.snopes.com/; https://www.politifact.com/}, and researchers have also explored automatic fact-checking methods~\cite{Ciampaglia2015-nu,hassan2017toward}. Researchers have investigated automatic detection through content style~\cite{Perez-Rosas2017-tt,Feng2012-ph,Zhou2020-vx}, user profile~\cite{shu2018understanding}, and information propagation~\cite{Shu2019-ig,ma2018rumor,Wu2015-nu}. Our work differs from this body of work in that we focus on the users to understand what and how their behavior changed after being exposed to misinformation.

Another angle to study misinformation is to understand its impact on users and society. Psychologists and computer scientists have studied the impact of misinformation by looking at changes in user's beliefs and the overall social network. Researchers have shown that continued exposure to unsubstantiated rumors makes users more credulous~\cite{Bessi2015-cc,Mocanu2015-hk}. Scientists have also shown that misbeliefs can persist after being exposed to misinformation~\cite{Nyhan2010-sf}. Loomba et al. focused on COVID vaccine related misinformation and found that users' vaccine intent decreased after exposure via qualitative analysis~\cite{Loomba2021-ai}. Dutta et al. looked at the role of the political campaign during the 2016 United States Presidential Election by Russia's Internet Research Agency (IRA) among Twitter users~\cite{Dutta2021-xo}. Researchers have investigated the correlation between misinformation and society polarization~\cite{Ribeiro2017-uk,Vicario2019-ov}.  Holme et al. also studied the influence on social network structure via simulations~\cite{holme2019impact}. In contrast, our work investigates the specific behavioral differences of users before and after the exposure to misinformation. To the best of our knowledge, this is the first work performing a large-scale, quantitative analysis on users' behavioral change after being exposed to a broad range of misinformation. 
\section{Methodology}
\subsection{Data Collection}
\subsubsection{Collecting Misinformation Tweets and the Exposed Users' Tweets} 
\begin{figure}[ht]
\caption{A PolitiFact article debunking a Facebook post}
\centering
\includegraphics[width=0.25\textwidth]{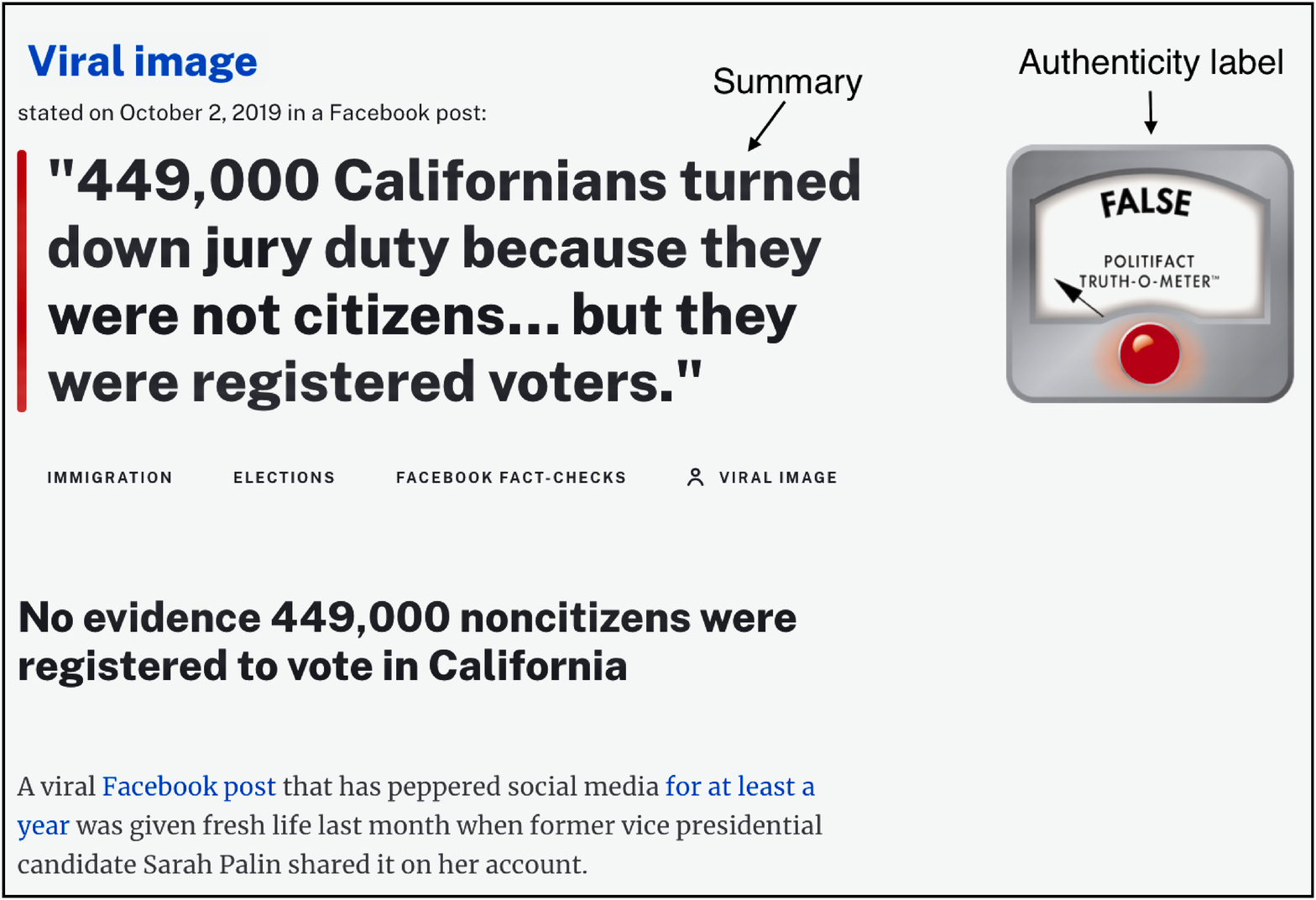}
\label{fig:PolitiFact}
\end{figure}

\begin{figure}[ht]
\caption{A sample tweet containing the misinformation}
\centering
\includegraphics[width=0.25\textwidth]{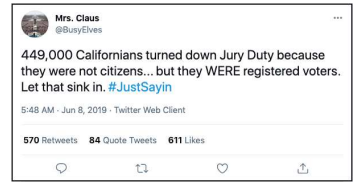}
\label{fig:tweet}
\end{figure}

The goal of this research is to understand behavioral changes of Twitter users after being exposed to misinformation tweets. Therefore, the first step is to identify the tweets that have misinformation content. As there is no ``gold-standard" misinformation detection model, we resorted to the expert fact-checked news source from PolitiFact as our ``seed" to find the corresponding tweets. Since PolitiFact does not work on tweets, we crawled all the fact-checked Facebook text-only and viral image posts which are labelled as ``pants on fire", ``false" or ``mostly false" from May 18, 2013 until Jan 31, 2021 (most of them are during 2018 to 2021). We crawled Facebook posts as it is also a social network platform and posts have a high probability of showing up on Twitter. Although PolitiFact also debunks politicians' and celebrities' claims, it turned out to be harder to directly search for them on Twitter. Figure~\ref{fig:PolitiFact} is an example of a fact-checked Facebook post on PolitiFact. For each debunked news post, we used the provided summary as the search term to search for the corresponding tweets on Twitter. To avoid unrelated results, we disregarded the posts whose summary was less than seven words. From the search response, we extracted the top-five tweets ranked by reply count. We removed the tweets that originated from fact-checking organizations, or that included any keyword regarding its veracity, e.g. ``conspiracy theory", ``debunk", and ``fake news". We crawled 1,119 debunked news posts from PolitiFact and we found 442 of them on Twitter. From the search results, we were able to collect 529 tweets with misinformation content. Figure~\ref{fig:tweet} shows an example of a tweet that we studied. After collecting all the tweets , we did a thorough verification to make sure our dataset only contained tweets with misinformed content. We manually removed all the tweets which were not misinformation and the users who replied to them, resulting in 399 tweets.

For each of the collected tweets, we identified all the users who replied to it, which resulted in 25,619 users. Since the before and after analysis was performed on the users who replied to the tweets, and each user replied at a different time, the ``zero" time for each user refers to the time of the user's first reply to the tweet. This method allows us to aggregate before and after behavior across different users who were exposed to misinformation tweets at different times. For each user, we collected all of their tweets starting at six months before their respective zero time and until six months after their respective zero time. A period of 30 days was used in place of a calendar month.

In order to ensure the users under the purview of this study were legitimate users, we used Botometer~\cite{Botometer} to remove users that were identified as potential bots. Botometer uses features from a user’s profile along with machine learning to identify accounts primarily run with the help of automation software, and will return a score representing the probability that an account is run by a bot. Users with score more than 0.5 were removed. There were 372 misinformation tweets with 21,071 replied users for target group after potential bot removal.

The long-term analysis includes only a subset of users who had activities throughout the whole 12-month period. The reason we had to exclude some users is that not all users had 6 months' of activity before or after the exposure. This way we only included the users for whom we had a complete 12 months of activity. This analysis also excludes users who joined Twitter within 6 months before the reply. There were 11,585 users for long-term analysis after the filtering.

\subsubsection{Generating the Baseline User Tweets Dataset}
In order to ensure that observed user behavior was sufficiently different from the general Twitter population, we built two user groups as baseline groups. Both groups were used as the baselines for the short-term analysis (24 hours) and one of them was used for the long-term analysis (6 months). For the first baseline group, we used the target group to understand the behavioral change when the same users were exposed to tweets without misinformation. To construct this baseline, we randomly collected 5 replies (exposures) to other tweets for each user and collected tweets before and after the exposure within the short term (24 hours). The analyses on this baseline show the average behavior of the 5 exposures. We selected 5 exposures because we were not able to confirm if other exposures were to true news tweets, so we averaged multiple exposures to eliminate it. This baseline is only used for the short-term analysis because there is a high possibility of overlapping periods for different exposures when looking at longer term behavior and it may interfere with the analysis. 

For the second baseline group, we collected tweets from a different set of users who were exposed to content similar in subject matter to the target group but also true in nature. Because it is difficult to find related tweets without misinformation, we searched the tweets that only contained true news. To achieve this, we extracted the entity from each misinformation tweet's text content using Open Information Extraction (OpenIE) tool of Stanford CoreNLP~\cite{coreNLP}. Then, we collected all the recent tweets from known true news sources~\cite{Horne2017-dw} and excluded some questionable ones in recent years (e.g. The Guardian) and did the same entity extraction process. For each entity from misinformation tweets, we chose a tweet with the same entity from the true-information tweet group with similar reply count. Due to the limitation of our crawling tools, only 3,200 most recent tweets could be fetched for each source, thus not all misinformation tweets could be matched. For the remaining unmatched misinformation tweets, we used their entities as the search term to search for related tweets. To ensure the tweet's veracity, we only considered tweets posted by verified accounts and gave priority to the known true news sources. We then selected tweets whose reply count was close to entity-matched misinformation tweets. Finally, we performed the same user scraping process as before to get users' tweets from 6 months before until 6 months after the exposure, and then removed potential bots. Table~\ref{tab:user_count} shows the actual number of users we considered for the analysis. There are far fewer users in baseline 2 group for the long-term analysis because many of the tweets collected from reliable sources were very recent ones and the exposed users did not have 6-months worth of activities after exposure. \\

Retweets and favorite tweets were not utilized within this work, as their presence could be unevenly distributed due to a time-sensitive constraint. We used Twint~\cite{Twint}, a Twitter-scraping Python library, to search and collect the tweets as described above. Twint can only retrieve the most recent retweets and favorite tweets. When working with non-recent data, it is unlikely that a majority of favorite tweets from said period will be reachable. Twint is also limited in the state of the accounts it is able to retrieve. It is unable to retrieve deleted account data, and tweets posted when the corresponding account is deleted or private. To ensure completeness and fairness of our dataset, we decided to exclude retweets and favorite tweets as well as accounts for which tweets could not be reached. 

\begin{table}[h]
\caption{Number of users for the analysis}
\centering
\scalebox{0.8}{\begin{tabular}{ |c|c|c| } 
\hline
User group & Analysis type & No. of users \\
\hline
\multirow{2}{*}{ \shortstack{Target group (same as baseline 1)}} & Short-term & 21,071 \\ 
\cline{2-3}
  & Long-term & 11,585 \\ 
\hline   
\multirow{2}{*}{Baseline 2}& Short-term & 19,357 \\
\cline{2-3}
  & Long-term & 5,970 \\
\hline
\end{tabular}}
\label{tab:user_count}
\vspace{-6pt}
\end{table}

\subsection{Features}
We analyzed the before and after user behavior through three specific metrics: average tweet count, average sentiment score, and language usage distance. All features were studied hourly (short-term) and monthly (long-term).

\textit{Tweet Count} was determined for each user by counting the total number of tweets posted by the user within the bounds of a 1-hour or 1-month period. This count includes tweets posted by the user and replies to other accounts during that time period. We excluded favorite tweets, retweets of tweets authored by another Twitter account, or other Twitter activity from the tweet count because of the limitations mentioned in the data collection section.

\textit{Sentiment Score} was calculated using VADER, a lexicon and rule-based sentiment analysis tool that is specifically attuned to sentiment expressed in social media~\cite{hutto2014vader}. A sentiment score within the range [-1, 1] was assigned to each tweet based on its content. A score close to -1 indicates a highly negative sentiment while a score close to 1 indicates a highly positive sentiment. These values were then averaged to form the hourly or monthly \textit{Sentiment Score} for each user.


\textit{Language Usage Distance} was used to evaluate the difference of language between two groups of tweets text. Similar to prior work~\cite{Hessel2016-il,Althoff2016-cg}, we adopted the Jensen-Shannon Divergence~\cite{fuglede2004jensen} to measure the unigram difference (hourly and monthly) in tweets as the language distance. A larger distance indicates a larger difference in language (word) usage. 
We removed all the mentions, URLs and stopwords from the tweets, and stemmed the words as the pre-processing step. 

We used dependent sample t-test to assess the statistical significance of the results for both \textit{Tweet Count} and \textit{Sentiment Score}. We used this type of test because the before and after samples are not independent of each other. We aggregated the users' hourly/monthly feature before and after the exposure to conduct the tests. 
\section{Results}
\subsection{RQ1: Do the users who reply to misinformation tweets exhibit a change in their behavior after the exposure?}
\begin{figure*}[htbp]
\centering
    \begin{subfigure}[b]{0.5\linewidth}
      \includegraphics[width=0.72\textwidth]{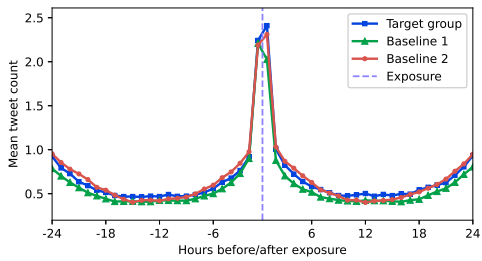}
      \caption{Short-term: hourly tweet count}
      \label{fig:rq1_count_short}
    \end{subfigure}\hfill
    \begin{subfigure}[b]{0.5\linewidth}
      \includegraphics[width=0.9\textwidth]{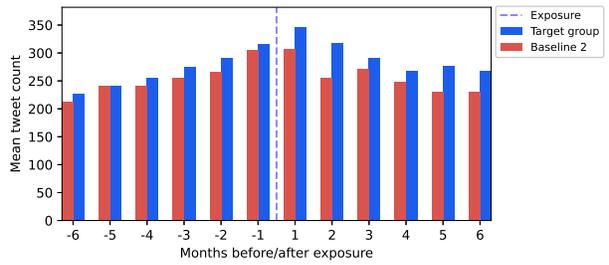}
      \caption{Long-term: monthly tweet count}
      \label{fig:rq1_count_long}
    \end{subfigure}
\caption{Average hourly (left) and monthly (right) tweet count}
\label{fig:rq1_count}
\end{figure*}

\textbf{The target group significantly increased tweeting frequency following their exposure to the misinformation tweets in both short and long term, compared with the baseline users.} As shown in Fig.~\ref{fig:rq1_count_short}, in the short-term analysis all three groups' tweeting frequency had a 24-hour periodic change. The target group's tweeting frequency increased significantly (0.66 vs. 0.68, P=5.7$\times10^{-12}$) during the 24-hour period after the exposure, while baseline 1 group's decreased significantly (0.60 vs. 0.59, P=4.1$\times10^{-5}$). 
For baseline 2 users, there is no significant tweeting frequency change (0.67 vs. 0.68, P=0.24). 
Note that we also analyzed the behavior for a 72-hour period and it showed the similar pattern. Due to space limitations, we only report the 24-hour analysis.

The long-term analysis had a similar pattern, as shown in Fig.~\ref{fig:rq1_count_long}. Although monthly tweet count increased for both groups, the target group's tweet count increased more significantly. The baseline 2 group's change was significant (253.5 vs. 257.6, P=0.013), but the change and significance level is much weaker than that of the target group (267,7 vs. 294.9, P=2.6$\times10^{-132}$).

\begin{figure*}[htbp]
\centering
    \begin{subfigure}[b]{0.5\linewidth}
      \includegraphics[width=0.72\textwidth]{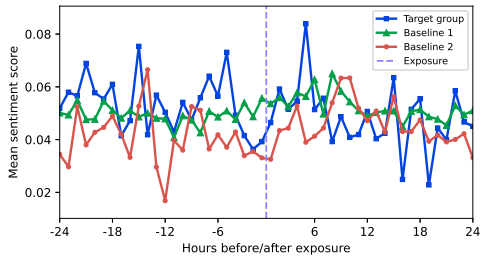}
      \caption{Short-term: hourly sentiment score}
      \label{fig:rq1_sent_short}
    \end{subfigure}\hfill
    \begin{subfigure}[b]{0.5\linewidth}
      \includegraphics[width=0.9\textwidth]{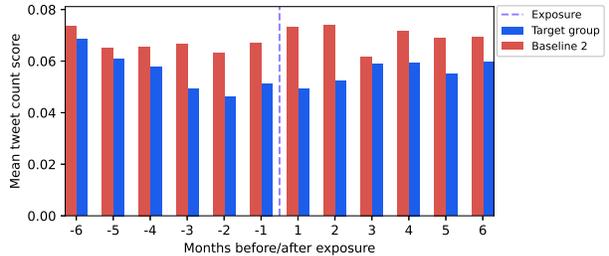}
      \caption{Long-term: monthly sentiment score}
      \label{fig:rq1_sent_long}
    \end{subfigure}
\caption{Average hourly (left) and monthly (right) sentiment score}
\label{fig:rq1_sent}
\end{figure*}

\textbf{The target group users did not change sentiment significantly in neither short nor long term.}  Sentiment score of all the 3 groups in the short-term did not change significantly (Fig.~\ref{fig:rq1_sent_short}). As shown in Fig.~\ref{fig:rq1_sent_long}, the target group's sentiment score didn't change significantly in long-term either (0.056 vs. 0.056, P=0.85), while baseline 2 group's sentiment score had a little increase (0.067 vs. 0.070, P=1.3$\times10^{-5}$). We argue that this is because a user's sentiment does not necessarily change in one direction (only increase or decrease) after the exposure. A person may express the same stance/opinion toward a tweet by using negative or positive language~\cite{Mohammad2017-ma,aldayel2019assessing}, which would not change the average sentiment score significantly.

\begin{figure}[htbp]
    \centering
    \begin{subfigure}[b]{0.8\linewidth}
      \includegraphics[width=\textwidth]{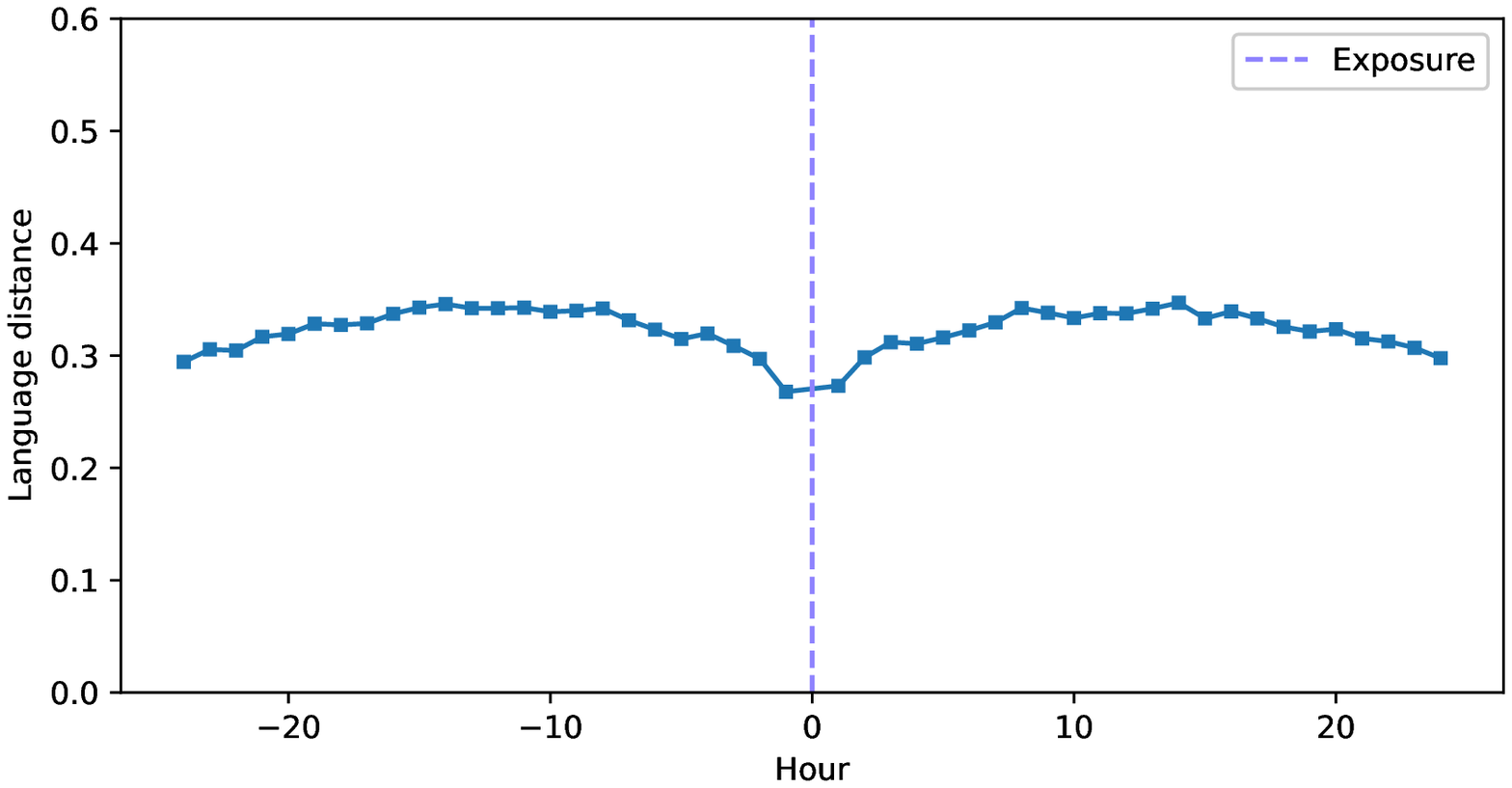}
      \caption{Short-term language distance}
    \end{subfigure}
    \begin{subfigure}[b]{0.8\linewidth}
      \includegraphics[width=\textwidth]{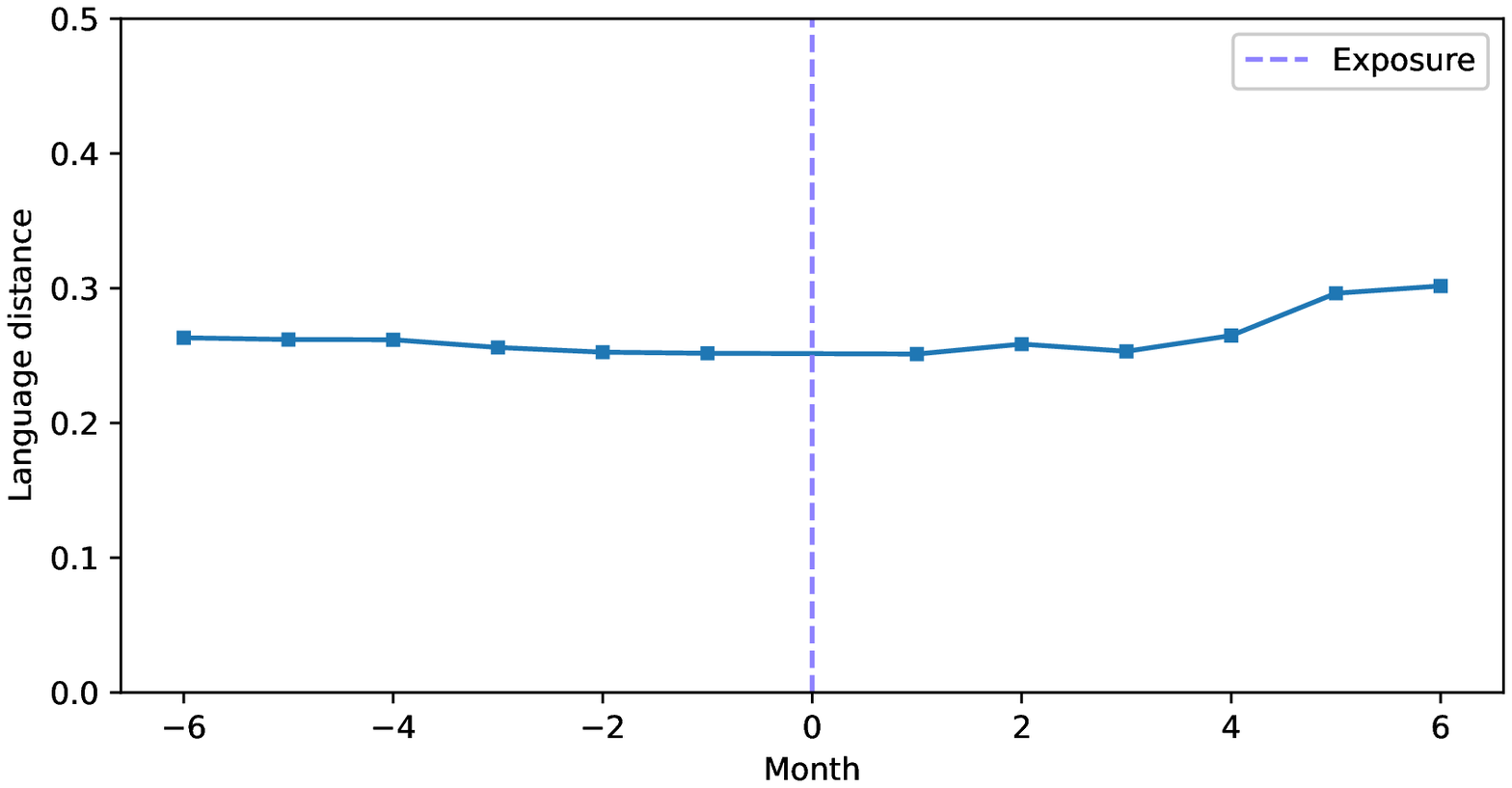}
      \caption{Long-term language distance}
    \end{subfigure}
\caption{Language distance between target and baseline 2 group}
\label{fig:lang_dis}
\end{figure}

To understand this further, we compared the target group with the baseline 2 group by their language distance. We calculated the language distance for each hour/month before and after the exposure between the target and the baseline 2 group. As shown in Fig.~\ref{fig:lang_dis}, the language distance of the target group with the baseline group is stable before and after the exposure, and there is a slight increase starting from the fourth month after exposure. This observation indicates that the target and the baseline 2 group users already have different language characteristics even before their respective exposure and that this difference does not change much after the exposure. This indicates that the misinformation and true information tweets attract users with different characteristics. 

\subsection{RQ2: Does the change in behavior of users who reply to {\it multiple} misinformation tweets differ from other users?}

\textbf{Multi-exposure users show a significant change of tweet count in long-term but not in short-term, and the change is weaker than that of other users.} We consider multi-exposure users to be the ones who replied to at least two misinformation tweets. Although users may reply to other misinformation tweets, in this work we only consider the exposure to our collected tweets. There are 504 users in this group. 

\begin{figure*}[htbp]
\centering
    \begin{subfigure}{0.5\linewidth}
      \includegraphics[width=0.72\textwidth]{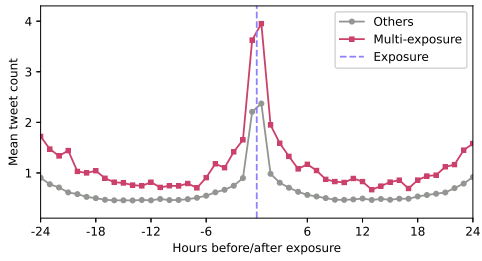}
      \caption{Short-term: hourly tweet count}
    \end{subfigure}\hfill
    \begin{subfigure}{0.5\linewidth}
      \includegraphics[width=0.9\textwidth]{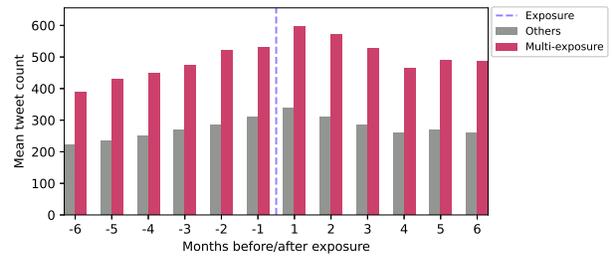}
      \caption{Long-term: monthly tweet count}
    \end{subfigure}
\caption{Average hourly (left) and monthly (right) tweet count for multi-exposure users.}
\label{fig:rq2_count}
\end{figure*}

As shown in Fig.~\ref{fig:rq2_count}, the tweet count for multi-exposure users did not have significant change in the short-term (1.14 vs. 1.17, P=0.37), while other users' (single-exposure) increase was statistically significant (0.64 vs. 0.67, P=7.3$\times10^{-12}$). In the long-term, multi-exposure users still had an increased tweet count (465.0 vs. 523.0, P=7.3$\times10^{-13}$), but its
significance level is weaker than that of the single-exposure users (261.2 vs. 287.3,  P=1.3$\times10^{-121}$). As mentioned in  RQ1’s result, we did not observe significant sentiment change .

\begin{figure*}[htbp]
\centering
    \begin{subfigure}{0.5\linewidth}
      \includegraphics[width=0.72\textwidth]{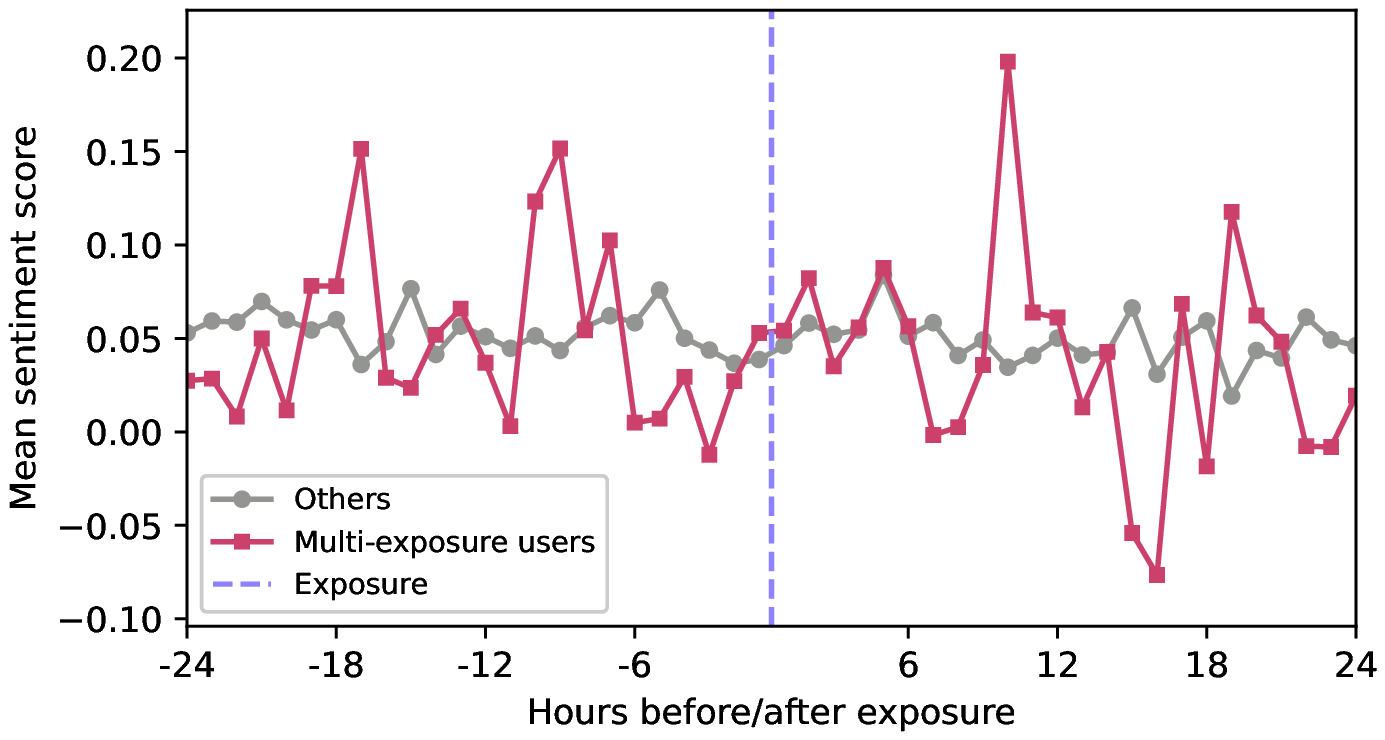}
      \caption{Short-term: hourly sentiment score}
      \label{fig:rq2_sent_short}
    \end{subfigure}\hfill
    \begin{subfigure}{0.5\linewidth}
      \includegraphics[width=0.9\textwidth]{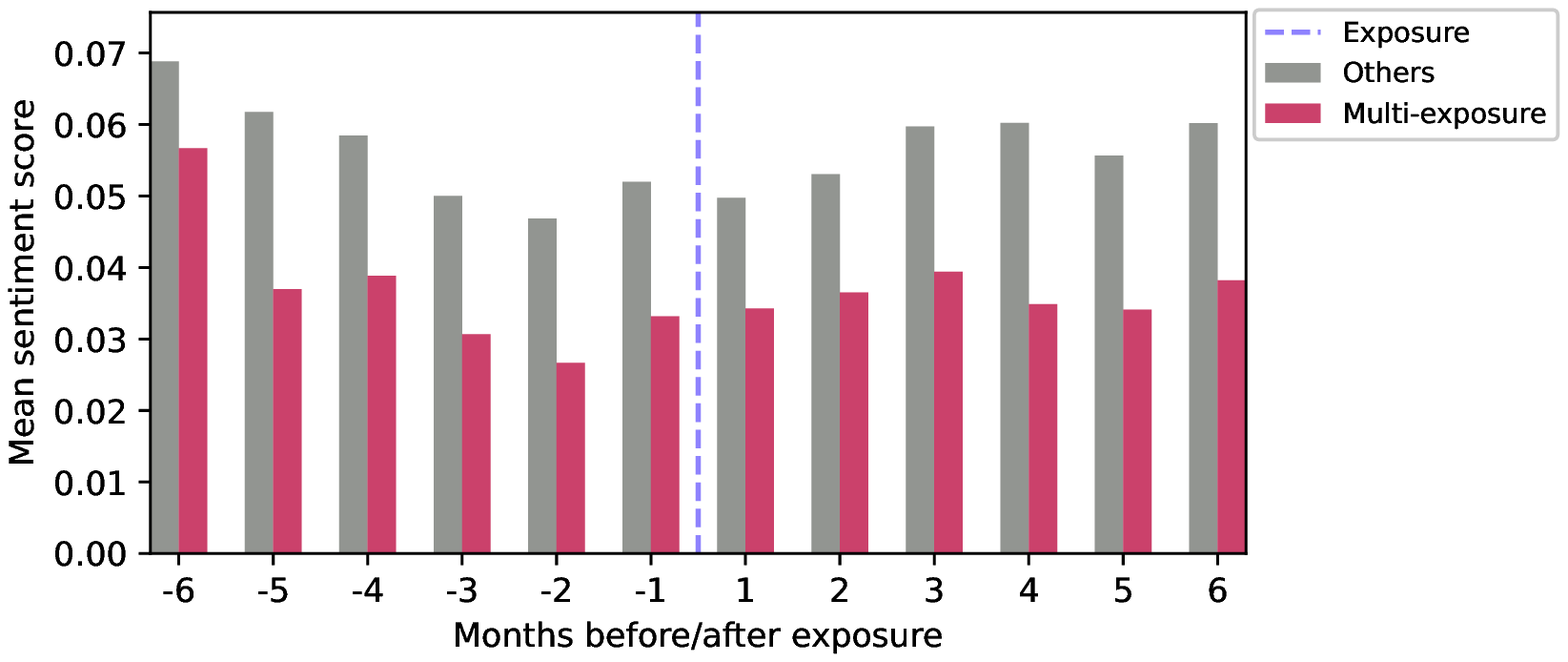}
      \caption{Long-term: monthly sentiment score}
      \label{fig:rq2_sent_long}
    \end{subfigure}
\caption{Average hourly (left) and monthly (right) sentiment score for multi-exposure users.}
\label{fig:rq2_sent}
\end{figure*}

From the comparison between the multi-exposure and single-exposure groups, it is shown that the multi-exposure group generally posts more tweets (Fig.~\ref{fig:rq2_count}) with more volatile sentiment (Fig.~\ref{fig:rq2_sent}), and this difference is stable across the 12 months (their long-term sentiment score is lower because averaging the volatile score in a longer term results in average monthly score closer to 0). We conclude that the multi-exposure users were already on a ``high-level mood" and their change was not as significant as that of the single-exposure users, who were rising from a relatively lower level.

\subsection{RQ3: What are the characteristics of the users who undergo extreme behavioral change after being exposed to misinformation tweets?}
Another interesting angle is to study the users who changed their behavior the most after the exposure. To separate this group of users, we calculate the tweet count increase and absolute sentiment score change from 1 hour/month before to 1 hour/month after the exposure. The users who had the top 5\% tweet count increase or sentiment change were selected, respectively. This resulted in 1,007 and 939 users who had extreme tweet count increase for the first hour/month, respectively. The the first hour/month extreme sentiment score change group had 1,054 and 941 users, respectively.

\textbf{Extreme tweet count increase and extreme sentiment score change do not align.} We first examined if the users having extreme tweet count increase and those having extreme sentiment change overlap. There were only 7 and 8 overlapped users for the first hour/month, respectively. This means that after being exposed to misinformation, ``furiously" posting tweets didn't occur together with sharp change of sentiment.

\begin{figure}[htbp]
    \centering
    \begin{subfigure}[b]{0.55\linewidth}
      \includegraphics[width=\textwidth]{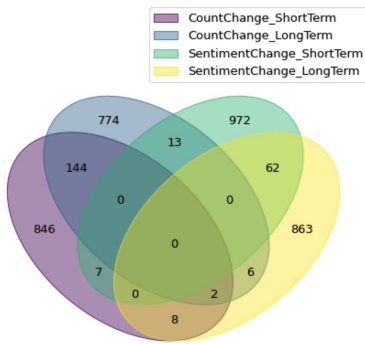}
    \end{subfigure}
\caption{Number of users in the extreme-change user groups. }
\label{fig:rq3}
\end{figure}

\textbf{Short-term and long-term change do not align.}
We also compared the tweeting frequency between the short and long-term among the same extreme-change users. Users who increased tweet count a lot in the first hour didn't show a large increase in the following months. Similarly, the users who increased tweeting frequency a lot in the first month didn't show the same level of increase in the first several hours. There were 146 and 62 overlapped users for extreme tweet count and sentiment change, respectively. Fig.~\ref{fig:rq3} visualizes the overlap across different extreme-change user groups.
\subsection{RQ4: Does the change in the behavior of the users after being exposed to misinformation tweets vary based on the number of their followers or the number of followers of the tweet authors?} 
\begin{figure}[htbp]
    \centering
    \begin{subfigure}[b]{0.75\linewidth}
      \includegraphics[width=\textwidth]{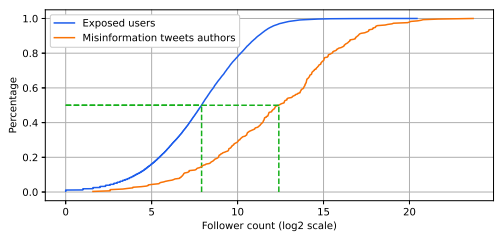}
    \end{subfigure}
\caption{CDF plot of follower count}
\label{fig:cdf_follower}
\end{figure}

We conducted two analyses for this research question, where the first is to understand if the exposed users behaved differently when their follower count is different, and the second is to understand if the exposed users behaved differently when the misinformation tweet authors' follower count is different. We separated the exposed users into low-follower count and high-follower count groups, where 240 was chosen to be the threshold for high and low follower count because 240 divided the users fairly well into two halves. Using the same idea for the misinformed tweets authors, 5400 was chosen as it separates the authors into two halves. Fig.\ref{fig:cdf_follower} shows the distribution of the followers. As a result, there were 13,797 and 9,325 users exposed to tweets authored by high-follower count users for short and long-term respectively, while there were 1,597 and 1,004 users exposed to tweets authored by users with low-follower count for short and long-term respectively. 

\textbf{Popular exposed users' (high-follower count) and less-popular users (low-follower count) both increased their tweeting frequency.} These two user groups didn't behave differently. The high-follower count users did have significant tweet count increases in short-term (1.03 vs. 1.05, P=0.006) and long-term (367.5 vs. 403.5, P=4.3$\times10^{-86}$). The low-follower count users were similar (0.56 vs. 0.60, P=4.3$\times10^{-12}$ for short-term, and 143.2 vs. 159.6, P=8.2$\times10^{-58}$ for long-term). As mentioned in RQ1's result, we did not observe significant sentiment change.

\begin{figure}[htbp]
    \centering
    \begin{subfigure}[b]{\linewidth}
      \includegraphics[width=\textwidth]{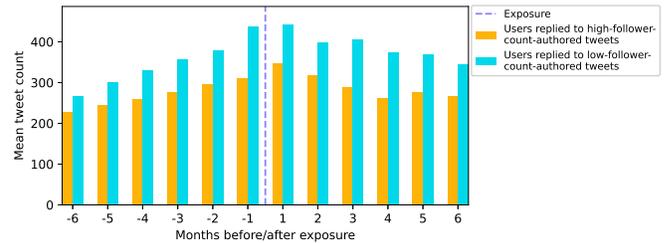}
    \end{subfigure}
\caption{Average hourly tweet count for exposed high-follower and low-follower count users}
\label{fig:rq4_sample}
\end{figure}

\textbf{Users exposed to high-follower-count-authored and low-follower-count-authored misinformation tweets both increased their tweeting frequency.} These two user groups didn't behave differently, either. Users exposed to high-follower authors' tweets had a significant increase in tweeting frequency for short term (0.78 vs. 0.80, P=1.03$\times10^{-6}$), and long term (269.6 vs. 293.3, P=8.1$\times10^{-85}$).  Users exposed to high-follower authors' tweets were similar (1.02 vs. 1.11, P=3.6$\times10^{-7}$ for short-term, and 345.7 vs. 389.4, P=9.1$\times10^{-17}$ for long-term). Fig.\ref{fig:rq4_sample} shows the long-term change and the users exposed to low-follower-count-authored tweets generally post more tweets. As mentioned in RQ1's result, we did not observe significant sentiment change.

\section{Concluding Discussion}
This paper investigates the behavior of Twitter users before and after being exposed (replied) to misinformation tweets. Our analysis reveals that users' tweet count significantly increases after exposure in both short and long-term. We do not find significant change in users' sentiment score. Through language distance analysis, we find that different user groups (target group and baseline group) are already different before their respective exposure. We also find that users who are exposed to more than one misinformation tweet have weaker changes than those who are only exposed to one, and find these multi-exposure users are already at a high-activity level before exposure. For users who have extreme behavioral changes, we find that their tweet count increases and sentiment score changes do not align. The short-term and long-term change do not align either. Another finding is that popular exposed users and less-popular users both increased their tweeting frequency, and this finding also holds for the users that are exposed to misinformation tweets authored by high-follower count users and low-follower count users, while the users exposed to low-follower-count-authored tweets generally post more tweets.

\textbf{Implications.} 
Our work reveals the positive correlation between users' tweeting frequency increase and exposure to misinformation tweets, which can potentially encourage more research investigating the misinformation's impact on specific user behaviors. Our work also has important implications for social platform designers and moderators. Misinformation does not affect all users equally and only a small number of users exhibit significant behavioral changes. Our second and third research questions give a closer look at these groups of interest.
We also find that the behavioral changes are similar for exposed users and misinformation tweets authors' with different follower count. These insights tell that all users could be potential target or disseminator of misinformation, which means the platform moderators should take care of all users when designing misinformation mitigation strategies.


\textbf{Limitations and Future Work.} 
Our work does not prove causality, i.e., while we observe significant changes in behavior before and after exposure to misinformation, we cannot definitively attribute this correlation to being primarily or even exclusively caused by the exposure. Although we built the baseline groups to eliminate some factors such as user personalities (baseline 1) and the entity of the tweet (baseline 2), there may be other unforeseen factors that cause these changes. For example, long-term tweet counts may also have risen because users spend more time on Twitter.

Another limitation lies with respect to the dataset. First, we only collected ``source" misinformation from Politifact, which is a small amount of misinformation and most of them are related to politics. A possible future direction is to collect more categories of misinformation (technology, business, etc) and study if changes in users' behavior are different for different types of misinformation. Second, due to Tweepy's limitation, this work is not able to access several critical sources of archival Twitter data including both user retweets and favorites during the necessary time period for the majority of the users. These interactions can be a good and important source to study users' attitude and preference after the exposure. As it takes different efforts to reply, retweet and favorite a tweet, a future direction is to expand the ``exposure" to retweets and favorites, and compare the differences of the behavior change of users with different types of exposure. Third, the baseline 1 was generated by averaging 5 other exposures because we didn't know if other exposed tweets are misinformation, which might cause the effect of "flattening" the behavioural changes. 

This work aims to study general misinformation's impact on users. When collecting data from Politifact, we didn't differentiate the authenticity levels labelled by the experts (pants on fire, false, mostly false). A future direction on this can be studying if users' behavior change differently after exposure to misinformation with different authenticity level. Furthermore, as it has been shown that there are also different strategies used by misinformation~\cite{Volkova2018-xx,Appling2015-lj}, understanding the effectiveness of different strategies and authenticity is important and useful for fighting misinformation, so that specific mitigation methods can be designed and applied accordingly. 

In addition, although this work has focused on ``first order" impact between the misinformation tweets and exposed users, this work may also raise the question of whether impacted users also impact their friends and followers through their retweets, replies and mentions, i.e., the ``second order" impact.



\bibliographystyle{IEEEtran}
\bibliography{IEEEabrv, biblio_rectifier}

\end{document}